\documentclass{elsart}
\usepackage{graphicx,amssymb}

\journal{Nucl. Instr. and Meth. in Phys. Res. A}

\begin{document}

\begin{frontmatter}

\title {Simulation of Three Dimensional Electrostatic Field Configuration
in Wire Chambers : A Novel Approach}

\author{N. Majumdar\corauthref{cor}},
\corauth[cor]{Corresponding author., Fax : 91 33 23374637}
\ead{nayana.majumdar@saha.ac.in}
\author{S. Mukhopadhyay}

\address{Nuclear Science Group, Saha Institute of Nuclear
Physics, 1/AF Bidhannagar, Kolkata - 700064, India}

\begin{abstract}
Three dimensional field configuration has been simulated for a simple wire
chamber consisting of one anode wire stretched along the axis of a grounded 
square cathode tube by solving numerically the boundary integral equation 
of the first kind. A closed form expression of potential 
due to charge distributed over flat rectangular surface
has been invoked in the solver using Green's function formalism leading to a
nearly exact computation of electrostatic field. 
The solver has been employed to study the 
effect of several geometrical attributes such as the aspect ratio
($\lambda = \frac{l}{d}$, 
defined as the ratio of the length $l$ of the tube to its width $d$)
and the wire
modeling on the field configuration. Detailed calculation has revealed 
that the field values deviate from the analytic estimates significantly when
the $\lambda$ is reduced to $2$ or below. The solver has
demonstrated the effect of wire modeling on the accuracy of the estimated
near-field values in the amplification region. The thin wire results can
be reproduced by the polygon model incorporating a modest number of surfaces 
($\geq 32$) in the calculation with an accuracy of more than $99\%$. The
smoothness in the three dimensional field calculation in
comparison to fluctuations produced by other methods has been observed.
\end{abstract}


\end{frontmatter}

\section{Introduction}
Wire chambers are often employed as tracking devices where it is necessary 
to detect and localize radiation. Starting from its application in nuclear 
and subnuclear physics, it has been employed in widely different fields 
such as biology, medicine, space, industrial radiology, over last three 
decades or more. The 
normal operation of a wire chamber is based on the collection of the 
charges created by direct ionization of the gas medium by the passage of 
radiation. The charges are collected on the electrodes by application of 
an electric field across the chamber. From the electric pulses, thus 
generated, the relevant information regarding the radiation is extracted. 
The flexibility in the design of wire chambers allows for
highly innovative and often considerably complex ones necessitating
meticulous investigations on their structure and performance.
The study of the electrostatic field plays a key role in optimizing the
design of these state of the art detectors to get a desired configuration 
for the field in a given volume as per the tracking requirement.
The analytic solution of the
field configuration for a specific geometry is always the best choice to 
do the same.
However, the analytic solution can be derived for severely restricted
geometries which is often not applicable to realistic and 
complicated wire chambers \cite{Erskine,Veenhof}. The diversity in 
the chamber design necessitates application of other techniques for 
numerical estimation like Finite Element Method (FEM) and Finite Difference 
Method (FDM) \cite{Buchanan,Lopez}. FEM is more widely used for the reason that it can seamlessly
handle any arbitrary geometry including even dielectrics. However, FEM has 
several drawbacks as well. It computes the potential at the nodes and 
the potential at non-nodal points can be obtained by interpolation only.
The inaccuracy generated by the interpolation technique can be made
arbitrarily small by proper meshing techniques at the cost of computation time
and efficiency. The more crucial aspect which harms the accuracy of the
estimation is the representation of the electric field by a low order, often
linear polynomial which is inadequate especially in the vicinity of the wires
where the field changes rapidly. The combination of inadequate representation
of the electric field and poor meshing lead to inaccurate estimation of the
field in the amplification region with the FEM technique.
The other approach which can yield
nominally exact result is Boundary Integral Equation (BIE) method. This method
is less popular due to its complicated mathematics and inaccuracies near the
boundaries. However, for the present problem of computation of electrostatic
field in wire chambers, BIE method is reasonably more suitable. It can
provide accurate estimate of the electrostatic field at any arbitrary
point by employing Green's function formulation which is necessary to
study the avalanche happening anywhere in the chamber due to the passage of 
radiation. A brief
comparison of BEM, the numerical implementation of BIE method, with FEM and 
FDM in the context of calculating three 
dimensional field configuration in wire chambers has been presented in 
\cite{IEEE}.

The major drawback of BEM is related to the approximations involved in
its numerical implementation. The approximations give 
rise to the infamous numerical boundary layer where
the method suffers from gross inaccuracies \cite{Renau}. This may lead to 
inaccurate
estimation of electrostatic field configuration which is not desirable in 
the close vicinity of the wires or the cathode. Recently, we have developed a 
novel approach in the 
formulation of BEM using analytic expressions for potential and
electrostatic field which leads to their nominally exact evaluation. The 
analytic expressions being valid throughout the physical volume, the 
formulation is capable of yielding accurate values even
in the near-field region. The application of this Nearly Exact Boundary
Element Method (NEBEM) solver \cite{ICCES05} for the very accurate estimation 
of 
electrostatic field in a wire chamber of elementary but useful geometry 
has been presented in this paper.

\section{Present Approach}
For electrostatic problems, the BIE can be expressed as
\begin{equation}
\label{eqn:BIE}
\phi(\vec r) = \int_S G(\vec r, \vec r^\prime) \rho(\vec r^\prime) dS^\prime
\end{equation}
where $\phi(\vec r)$ represents potential at $\vec r$ integrating the
integrand over boundary surface $S$, $\rho(\vec r^\prime)$ the
charge density at $\vec r^\prime$ and $G(\vec r, \vec r^\prime) = 
1/4\pi\epsilon |\vec r - \vec r^\prime|$ with $\epsilon$ being the
permittivity of the medium. The BIE is numerically solved by discretizing 
the charge carrying surface $S$ in a number of segments on which uniform
charge densities $\rho$ are assumed to be distributed. The discretization 
leads to a matrix representation of the BIE as follows
\begin{equation}
\mathbf{A} \cdot \mathbf{\rho} = \mathbf{\phi}
\label{eqn:Matrix}
\end{equation}
where $A_{ij}$ of $\mathbf{A}$ represents the potential at the mid-point of
segment $i$ due to a unit charge density distribution at the segment $j$.
For known potential $\mathbf{\phi}$, the unknown charge distribution 
$\mathbf{\rho}$ is estimated
by solving Eqn.(\ref{eqn:Matrix}) with the elements of influence matrix 
$\mathbf{A}$ modeled by a sum of known basis functions with constant unknown
coefficients.  

\begin{figure}[hbt]
\begin{center}
\includegraphics[height=2in,width=3in]{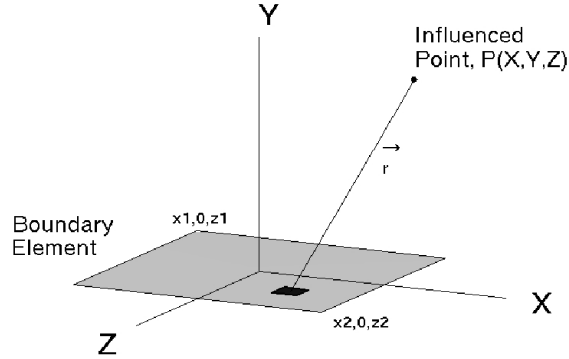}
\caption{A rectangular surface with uniform distributed source}
\label{fig:GeomElem}
\end{center}
\end{figure}

In the present approach, namely NEBEM, the influences are calculated 
using analytic solution of potential and electrostatic field due to a 
uniform charge distribution over a flat rectangular surface. The 
expression for the potential $\phi$ at a point $P(X,Y,Z)$ in free space due to
uniform unit charge density distributed on a rectangular surface having
corners at $(x_1,0,z_1)$ and $(x_2,0,z_2)$ as shown in Fig.\ref{fig:GeomElem}
can be represented as a multiple of
\begin{equation}
\label{eqn:Nebem}
\phi(X,Y,Z) = \int_{z_1}^{z_2} \int_{x_1}^{x_2} \frac{dx\,dz}{\sqrt{(X-x)^2 +
Y^2 + (Z-z)^2}}
\end{equation}
where the multiple depends upon the strength of the source and other physical
considerations. The closed form expression for $\phi(X,Y,Z)$ can be deduced
from the Eqn.(\ref{eqn:Nebem}).
This can be expressed as follows.
\begin{eqnarray}
\label{eqn:PotExact}
\lefteqn{\phi(X,Y,Z) =} \nonumber \\
&& (X-x_1)\,\ln \left( \frac{D_{12}\, -\, (Z-z_2)} {D_{11}\, -\, (Z-z_1)} \right) + (X-x_2)\,\ln \left( \frac{D_{21}\, -\, (Z-z_1)} {D_{22}\, -\, (Z-z_2)} \right) \nonumber \\
&& + (Z-z_1)\,\ln \left( \frac{D_{21}\, -\, (X-x_2)} {D_{11}\, -\, (X-x_1)} \right) + (Z-z_2)\,\ln \left( \frac{D_{12}\, -\, (X-x_1)} {D_{22}\, -\, (X-x_2)} \right) \nonumber \\
&& + \frac{i\, |Y|}{2} \nonumber \\
&& \left( \right. S_1\, \left( \right.
tanh^{-1} \left( \frac {R_1 + i\, I_1} {D_{11}\, \left| Z - z_1 \right|}  \right)
-\, tanh^{-1} \left( \frac {R_1 - i\, I_1} {D_{11}\, \left| Z - z_1 \right|}  \right) \nonumber \\
&& +\, tanh^{-1} \left( \frac {R_1 - i\, I_2} {D_{21}\, \left| Z - z_1 \right|}  \right)
-\, tanh^{-1} \left( \frac {R_1 + i\, I_2} {D_{21}\, \left| Z - z_1 \right|}  \right) \left. \right) \nonumber \\
&& +\, S_2\, \left( \right.
tanh^{-1} \left( \frac {R_2 + i\, I_2} {D_{22}\, \left| Z - z_2 \right|}  \right)\,
-\, tanh^{-1} \left( \frac {R_2 - i\, I_2} {D_{22}\, \left| Z - z_2 \right|}  \right) \nonumber \\
&& +\, tanh^{-1} \left( \frac {R_2 + i\, I_1} {D_{12}\, \left| Z - z_2 \right|}  \right)\,
-\, tanh^{-1} \left( \frac {R_2 - i\, I_1} {D_{12}\, \left| Z - z_2 \right|}  \right) \left. \right) \left. \right) \nonumber \\
&& - 2\,\pi\,Y
\end{eqnarray}
where
\begin{eqnarray*}
D_{11} = \sqrt { (X-x_1)^2 + Y^2 + (Z-z_1)^2 };
D_{12} = \sqrt { (X-x_1)^2 + Y^2 + (Z-z_2)^2 } \\
D_{21} = \sqrt { (X-x_2)^2 + Y^2 + (Z-z_1)^2 };
D_{22} = \sqrt { (X-x_2)^2 + Y^2 + (Z-z_2)^2 } \\
R_1 = Y^2 + (Z-z_1)^2;
R_2 = Y^2 + (Z-z_2)^2 \\
I_1 = (X-x_1)\,\left| Y \right|;
I_2 = (X-x_2)\,\left| Y \right|;
S_1 = {\it sign} (z_1-Z);
S_2 =  {\it sign} (z_2-Z)
\end{eqnarray*}

The electrostatic field can similarly be represented as a multiple of
\begin{equation}
\label{eqn:EFInt}
\vec{F}(X,Y,Z) = \int_{z_1}^{z_2} \int_{x_1}^{x_2}
            \frac{\hat{r}\,dx\,dz}{r^2}
\end{equation}
where $\vec{r}$ is the displacement vector from an infinitesimal area of the
element to the point
$P(X,Y,Z)$ where the field will be evaluated. The integration of Eqn.
(\ref{eqn:EFInt}) gives the exact expressions for the field in $X$, $Y$ and
$Z$-directions as follow.
\begin{equation}
\label{eqn:FxExact}
F_x(X,Y,Z) =
\ln \left( \frac{D_{11}\, -\, (Z-z_1)} {D_{12}\, -\, (Z-z_2)} \right) \,+\, \ln \left( \frac{D_{22}\, -\, (Z-z_2)} {D_{21}\, -\, (Z-z_1)} \right)
\end{equation}
\begin{eqnarray}
\label{eqn:FyExact}
\lefteqn{F_y(X,Y,Z) =} \nonumber \\
&& -\, \frac{i}{2}\, Sign(Y) \nonumber \\
&& \left( \right. S_1\, \left( \right.
tanh^{-1} \left( \frac {R_1 + i\, I_1} {D_{11}\, \left| Z - z_1 \right|}  \right)
-\, tanh^{-1} \left( \frac {R_1 - i\, I_1} {D_{11}\, \left| Z - z_1 \right|}  \right) \nonumber \\
&& +\, tanh^{-1} \left( \frac {R_1 - i\, I_2} {D_{21}\, \left| Z - z_1 \right|}  \right)
-\, tanh^{-1} \left( \frac {R_1 + i\, I_2} {D_{21}\, \left| Z - z_1 \right|}  \right) \left. \right) \nonumber \\
&& +\, S_2\, \left( \right.
tanh^{-1} \left( \frac {R_2 + i\, I_2} {D_{22}\, \left| Z - z_2 \right|}  \right)\,
-\, tanh^{-1} \left( \frac {R_2 - i\, I_2} {D_{22}\, \left| Z - z_2 \right|}  \right) \nonumber \\
&& +\, tanh^{-1} \left( \frac {R_2 + i\, I_1} {D_{12}\, \left| Z - z_2 \right|}  \right)\,
-\, tanh^{-1} \left( \frac {R_2 - i\, I_1} {D_{12}\, \left| Z - z_2 \right|}  \right) \left. \right) \left. \right) \nonumber \\
&& +\, \it{C}
\end{eqnarray}
\begin{equation}
\label{eqn:FzExact}
F_z(X,Y,Z) =
\ln \left( \frac{D_{11}\, -\, (X-x_1)} {D_{21}\, -\, (X-x_2)} \right) \, + \,\ln \left( \frac{D_{22}\, -\, (X-x_2)} {D_{12}\, -\, (X-x_1)} \right)
\end{equation}
In Eqn.(\ref{eqn:FyExact}), $C$ is a constant of integration as follows:
\[\it{C} = \left\{
\begin{array}{l l}
  0 & \quad \mbox{if outside the extent of the flat surface}\\
  2\, \pi & \quad \mbox{if inside the extent of the surface and Y $>$ 0}\\
  -2\, \pi & \quad \mbox{if inside the extent of the surface and Y $<$ 0} \end{array} \right. \]

All these equations 
have been used as foundation of the three dimensional solver \cite{EABE}.

In the present problem, two different modeling schemes of the wire have been
used to study the field configuration. When the wire has been modeled as a
polygon, the above expressions from Eqn.(\ref{eqn:PotExact})-
Eqn.(\ref{eqn:FzExact}) have been employed to estimate the potential and
the electrostatic field. In the other model, the wire has been considered
as a thin wire
where the radius of the wire $a$ has been assumed to be small compared to the 
distance $r$ of the observation point ($a << r$).
The expression for the potential at any point due to 
a wire element along $Z$-axis is the following.
\begin{equation}
\phi(X,Y,Z)\, =
2\, \pi\, a\, log
\left(
\frac{\sqrt{X^2 + Y^2 + (h+Z)^2} + (h+Z)}{\sqrt{X^2 + Y^2 + (h-Z)^2} - (h-Z)}
\right)
\end{equation}
where $h$ is the half of the length of the wire element. It should
be mentioned here that the analytic solution of the two dimensional 
electrostatic field
of a doubly periodic wire array in the Garfield code \cite{Garfield} is 
derived using a similar thin-wire approximation \cite{Erskine}.
The expressions for the electrostatic field components can be 
presented as the following under the same assumption.
\begin{equation}
F_x(X,Y,Z) =
2\, \pi\, a\, X\,
{ \left(
\frac
{ (h-Z) \sqrt {{X}^{2}+{Y}^{2}+{(h+Z)}^{2}}\, +\, (h+Z) \sqrt {{X}^{2}+{Y}^{2}+{(h-Z)}^{2}} }
{ \left( {X}^{2}+{Y}^{2} \right) \sqrt {{X}^{2}+{Y}^{2}+{(h-Z)}^{2}}\sqrt {{X}^{2}+{Y}^{2}+{(h+Z)}^{2}}}
\right) }
\end{equation}
\begin{equation}
F_y(X,Y,Z) =
2\, \pi\, a\, Y\,
{ \left(
\frac
{(h-Z) \sqrt {{X}^{2}+{Y}^{2}+{(h+Z)}^{2}}\, +\, (h+Z) \sqrt {{X}^{2}+{Y}^{2}+{(h-Z)}^{2}}}
{ \left( {X}^{2}+{Y}^{2} \right) \sqrt {{X}^{2}+{Y}^{2}+{(h-Z)}^{2}}\sqrt {{X}^{2}+{Y}^{2}+{(h+Z)}^{2}}}
\right) }
\end{equation}
\begin{equation}
F_z(X,Y,Z) =
2\, \pi\, a\,
{ \left(
\frac
{ \sqrt {{X}^{2}+{Y}^{2}+{(h+Z)}^{2}}-\sqrt {{X}^{2}+{Y}^{2}+{(h-Z)}^{2}}}
{ \sqrt {{X}^{2}+{Y}^{2}+{(h+Z)}^{2}} \, \sqrt {{X}^{2}+{Y}^{2}+{(h-Z)}^{2}}}
\right) }
\end{equation}
However, a separate set of expressions is needed to evaluate the potential and
electrostatic field due to a wire element along
its axis. These incorporate the effect of finite radius of the wire element and
are expressed below.
\begin{equation}
\phi(0,0,Z) =
2\, \pi \, a\, log
\left(
\frac{\sqrt{a^2+(h+Z)^2} + (h+Z)}{\sqrt{a^2+(h-Z)^2} - (h-Z)}
\right)
\end{equation}
In this case, only the $Z$-component of the field is non-zero and can be
written as
\begin{equation}
F_z(0,0,Z) =
2\, \pi\, a\,
\left(
{\frac {\left( \sqrt {{(h+Z)}^{2}+{{a}}^{2}} - \sqrt {{(h-Z)}^{2}+{{a}}^{2}} \right) }
{\sqrt {{(h-Z)}^{2}+{{a}}^{2}} \sqrt {{(h+Z)}^{2}+{{a}}^{2}}}}
\right)
\end{equation}

\section{Numerical Implementation}
\begin{figure}[htb]
\begin{center}
\includegraphics[height=2in,width=3in]{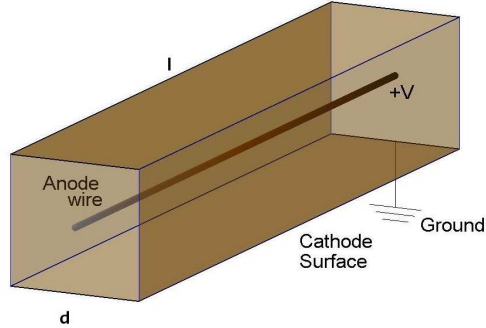}
\caption{Schematic representation of the wire chamber. The length and the
width of the square tube are represented by $l$ and $d$ respectively. The
anode wire along its axis has diameter $2a$. The wire is supplied a voltage
$+V$ and the cathode is kept grounded.}
\label{fig:Geometry}
\end{center}
\end{figure}

The present problem studied with the NEBEM is to compute the electrostatic 
potential and field for a simple geometry consisting of a single anode 
wire running along the
axis of a square tube. Similar configuration is used in Iarocci Tube, Limited
Streamer Tube etc. which are widely employed in various high energy physics
experiments \cite{Babar,Chorus}. It should be noted that no end plate has been 
considered in the model. A schematic diagram of the wire chamber has been 
illustrated
in Fig.\ref{fig:Geometry}. The anode wire has been supplied a positive high
voltage of $1000$ Volt and the surrounding cathode tube is grounded. 
Several cases for altered tube cross section ($d \times d$), aspect
ratio ($\lambda = \frac{l}{d}$) as well as two different wire models
have been studied. 
It should be noted here that if only the mid-plane estimates of the 
wire chamber are of importance, the computation time can be reduced 
drastically by using even one element in the axial direction resulting
into less than $100$ slender elements in total for the present problem.
This has been the case when the computation has been carried out for
the mid-plane properties of large aspect ratio chambers. On the other
hand, for proper three dimensional computation, the
four flat rectangular 
surfaces have been segmented in to $21$ elements along
the $X$-direction and $21$ in $Z$-direction. 
The anode wire when considered as a polygon has been modeled with
$32$ surfaces. The size of influence matrix has  
varied from $85 \times 85$ to $2436 \times 2436$ depending upon the
scheme of segmentation.

\section{Results}
\begin{figure}[htb]
\begin{center}
\includegraphics[height=3in,width=4.5in]{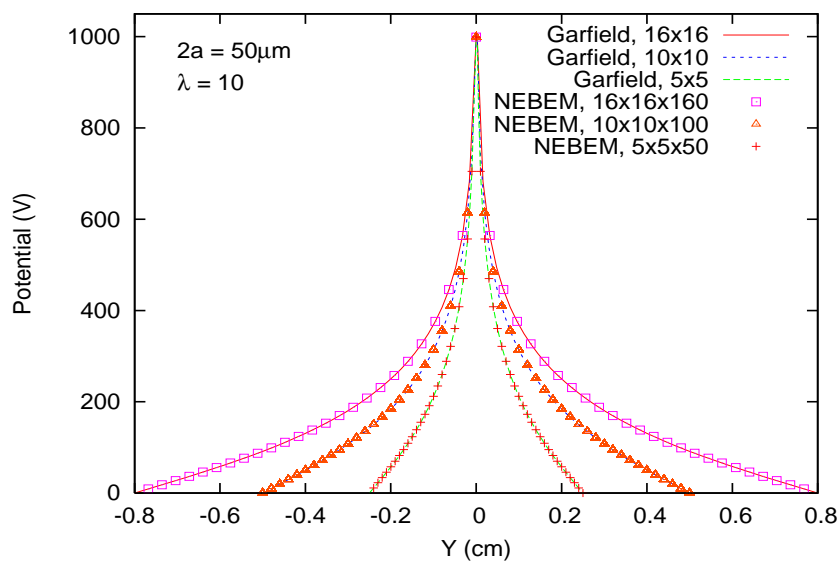}
\caption{Comparison of potential at the mid-plane of the chamber with
aspect ratio $10$ and wire diameter $50 \mu $m. Three variations in the
chamber cross-section are illustrated along with analytic values.}
\label{fig:Potiarocci}
\end{center}
\end{figure}

\begin{figure}[htb]
\begin{center}
\includegraphics[height=3in,width=4.5in]{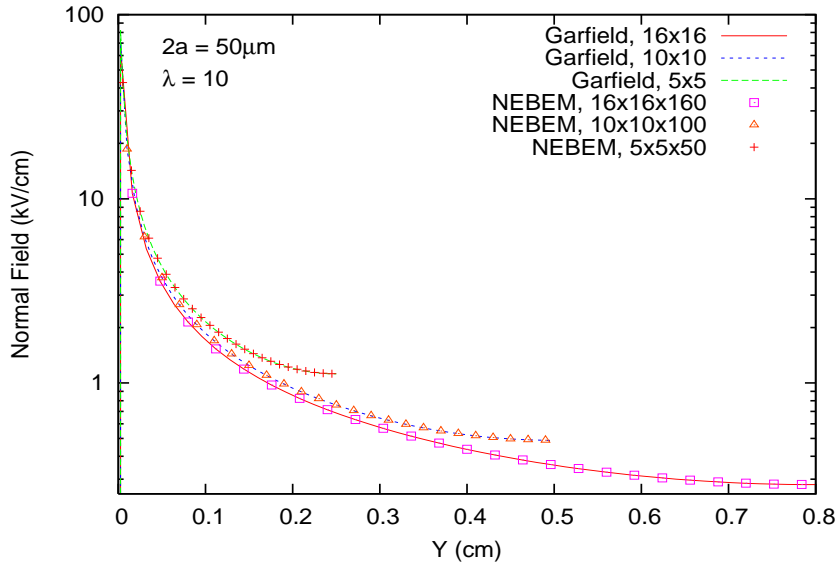}
\caption{Comparison of normal electric field at the mid-plane of the chamber 
with aspect ratio $10$ and wire diameter $50 \mu $m. Three cases of 
varied cross-sections are illustrated along with analytic estimates.}
\label{fig:EFiarocci}
\end{center}
\end{figure}

The NEBEM calculations for potential and normal electrostatic field 
($Y$-component) at the mid-plane of the chamber have been compared 
with the analytic estimates of an infinitely long tube provided by
the Garfield code \cite{Garfield} to demonstrate the accuracy of the 
solver. In Fig.\ref{fig:Potiarocci} and Fig.\ref{fig:EFiarocci}, 
the results are shown for a variation in the tube cross-section
from $5$mm
$\times 5$mm to $16$mm $\times 16$mm with wire diameter $50 \mu $m,
the wire being modeled as a polygon with $32$ surfaces. The aspect
ratio, $\lambda = \frac{l}{d}$, 
has been kept $10$ to retain the property of 
infiniteness so as to compare with analytic estimates of an infinitely
long tube. The comparison of two calculations with a spatial frequency
of $100 \mu $m shows an excellent agreement over the
whole range of tube dimensions. The NEBEM results calculated with
thin-wire approximation has not been included in these figures which
also yield similar agreement with the analytic ones.

\begin{figure}[htb]
\begin{center}
\includegraphics[height=3in,width=4.5in]{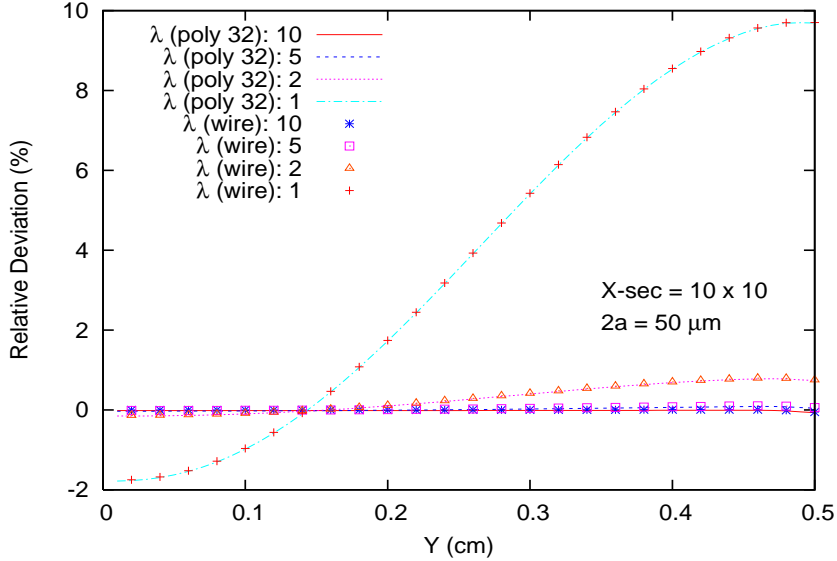}
\caption{Relative deviation of normal electric field from the analytic 
values at the mid-plane of the chamber
with varied aspect ratios for polygon and thin-wire modeling of the wire. 
The cross-section of the chamber and the 
diameter of the wire are $10$mm $\times 10$mm and $50 \mu $m respectively.}
\label{fig:Error}
\end{center}
\end{figure}

The difference of the NEBEM calculations from the analytic values have been
estimated as follows.
\begin{equation}
Relative Deviation (\%) = \frac{Garfield - NEBEM}{Garfield} \times 100
\label{eqn:error}
\end{equation}
This has been illustrated in Fig.\ref{fig:Error} by plotting the
relative deviation of NEBEM normal electrostatic field from the analytic
values calculated at the mid plane of the chamber. The relative deviations
estimated with thin-wire approximation have been plotted as well.
Since the NEBEM is a full-fledged three dimensional solver,
the effect of $\lambda$ of the tube on the field configuration can be 
studied using it. Several such estimates of relative deviations for
different aspect ratios have been shown in Fig.\ref{fig:Error} calculated
using both of polygon with $32$ surfaces and thin-wire models. 
The cross-section of the tube has been considered to be $10$mm 
$\times 10$mm with wire diameter $50 \mu $m.
It has been observed that the departure from the analytic solutions 
for an infinitely long tube becomes significant when $\lambda$ is reduced
to $2$ and below. It becomes apparent (close to
$1\%$) as $\lambda$ is dropped down to $2$ and enhances up to
$10\%$ when $\lambda$ is still reduced to $1$. The amount of relative 
deviation in the vicinity of the anode wire is maximum $2\%$ for the smallest
aspect ratio. The
trend is similar in both of polygon and thin-wire models as can be seen in the
figure.
It should be noted here that the use of end plates is expected to alter the
relative deviation particularly at smaller aspect ratios.

\begin{figure}[htb]
\begin{center}
\includegraphics[height=3in,width=4.5in]{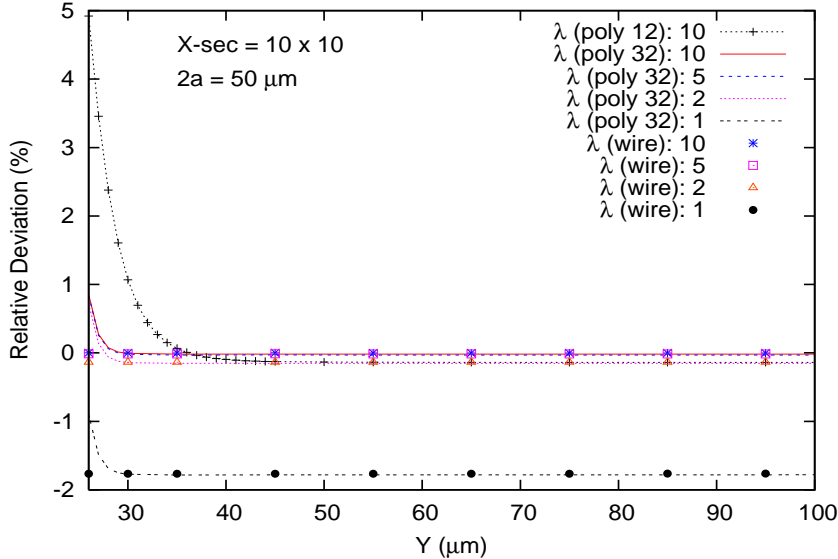}
\caption{Relative deviation in normal field from the analytic values at
close proximity to the anode wire. The tube cross-section and wire diameter
are $10$mm $\times 10$mm and $50 \mu $m respectively.}
\label{fig:Error_ex}
\end{center}
\end{figure}

The most essential study in such wire chambers is the field configuration 
in the amplification region which matters most in their performance. 
Since NEBEM can evaluate three dimensional field at any point in the physical
volume including the near-field region, a thorough study of the field values 
in the amplification region can be made using it. A comparative study has been 
carried out within twice the diameter from the wire-axis (i.e. $100 \mu $m),
the closest limit being
just $1 \mu $m away from the surface of the wire (i.e. $26 \mu $m) using
two different wire models. 
The calculations have been shown in Fig.\ref{fig:Error_ex} for the cases 
illustrated in Fig.\ref{fig:Error}.
Although the agreement between the polygon and thin-wire model is 
excellent up to quite close proximity of the wire, a departure has been 
observed within one radius to the wire
in case of polygon modeling. It has been observed that the departure
is almost negligible (below $1\%$) when larger number of surfaces (about $32$)
has been incorporated. It can increase to as high as $5\%$ when less number
of surfaces like $12$ is used. It is obvious from the calculation
that the thin-wire approximation is adequate to estimate the field
configuration in the near-field region in symmetric 
configurations. However, depending upon the nature of
the problem, the polygon model may be useful in calculation of azimuthal
variation of properties in an asymmetric configuration. In that case, 
a modest number of the 
polygon surfaces should be enough to obtain the field configuration
with high accuracy. 

\begin{figure}[htb]
\begin{center}
\includegraphics[height=3in,width=4.5in]{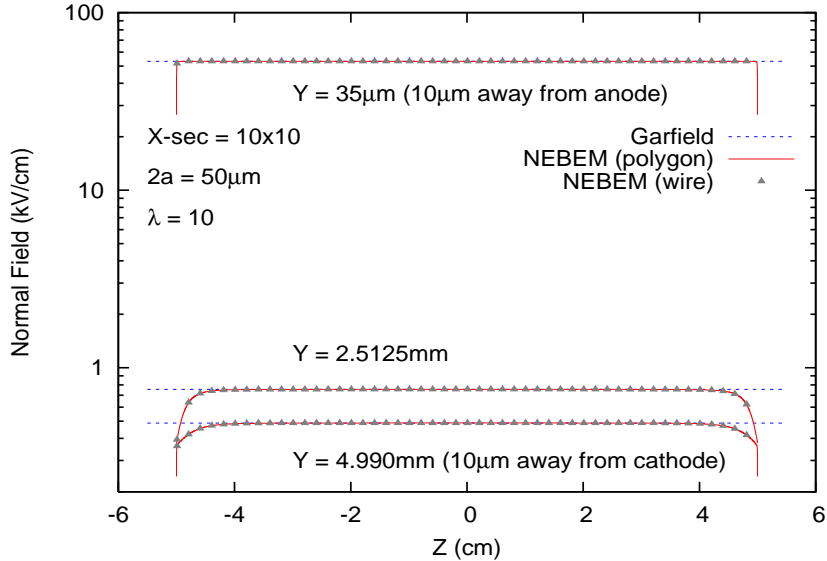}
\caption{Axial deviation of normal electric field at the mid-plane of the 
chamber with cross-section $10$mm $\times 10$mm, aspect ratio $10$ and wire
diameter $50 \mu $m, calculated at three $Y$-positions. Two different wire
models are considered.}
\label{fig:AxialEF}
\end{center}
\end{figure}

Finally, the variation of normal electrostatic field along the axial 
direction of the tube has been studied which has been plotted in 
Fig.\ref{fig:AxialEF}. The tube dimension has been considered to be
$10$mm $\times 10$mm $\times 100$mm with wire diameter $50 \mu $m. 
The calculations have been carried out at three different transverse
locations as indicated in the figure. The middle line represents the 
calculation done at halfway between the anode and the cathode. 
The two dimensional 
analytic solutions provided by the Garfield code have been illustrated 
in three dimension by the lines 
representing the uniform field configuration throughout the length. The NEBEM
results reproduce the two dimensional analytic values for more 
than $85\%$ of
the tube length. However, in the remaining $15\%$ towards the ends, the three 
dimensional effects are non-negligible. Even more important point to be noted 
here is that the NEBEM calculation produces perfectly smooth variation of the 
field with a spatial frequency of $10 \mu $m only while significant 
fluctuations are known to be present in FDM, FEM and 
usual BEM solvers because of their strong dependence on nodal properties.
This remarkable feature of the present solver should allow more
realistic estimation of the electrostatic field of various gas detectors
resulting into better gain estimations.

\section{Conclusion}
The three dimensional NEBEM solver has yielded accurate electrostatic
field configuration of a square tube wire chamber which represents the 
analytic estimates quite well in most of the detector volume when the
aspect ratio is large enough ($\lambda > 5$) except at the ends of
the chamber where end effects can be observed.
For smaller aspect ratios ($\lambda < 2$),  
non-negligible departures (about $2\%$) from the 
analytic values estimated for infinitely long chamber have been 
observed even in the amplification region. A large deviation (about 
$10\%$) has also been observed near the cathode surface.
The near-field calculation in the close vicinity to the
anode wire (within one diameter) has produced a 
difference in the results obtained with polygon and thin-wire models.
The observation has implied that in order to obtain accurate field
estimates with polygon modeling in the asymmetric configuration, 
an adequate number (e.g.$32$ for error $< 1\%$) of polygon surfaces 
are required to reproduce thin-wire results.
The simple but robust formulation of the solver
using closed form expressions can also be used to solve for gas detectors
of other geometries. Since the solver can produce very smooth
and precise estimate of three dimensional electrostatic field even in the
near-field region, it should
be very useful in providing important information related to the design and
interpretation aspects of a wire chamber. 

\section{Acknowledgement}
The authors are thankful to Prof. B. Sinha, Director, SINP, and Prof.
S. Bhattacharya, Head, NAP Division of SINP for their encouragement and
support throughout this work.

\end{document}